\documentclass{PoS}

\RequirePackage{xspace}
\usepackage{relsize}
\usepackage{url}
\usepackage{lineno}

\def\eg{\ensuremath {e.g.}\xspace}

\def\babar {{\mbox{\slshape B\kern-0.1em{\smaller A}\kern-0.1em B\kern-0.1em{\smaller A\kern-0.2em R}}}\xspace}

\def\superb {{Super\kern-0.04em\textit{B}}\xspace}
\let\SuperB=\superb
\def\CP        {\ensuremath{C\!P}\xspace}
\def\invab   {\ensuremath{\mbox{\,ab}^{-1}}\xspace}

\def\cms     {\ensuremath{{\rm \,cm}^{-2} {\rm s}^{-1}}\xspace}
\def\epem   {\ensuremath{e^+e^-}\xspace}
\def\nunub  {\ensuremath{\nu{\overline{\nu}}}\xspace}
\def\KL    {\ensuremath{K^0_{\scriptscriptstyle L}}\xspace}
\newcommand{\lum} {{\ensuremath{\cal{L}}\xspace}}
\def\FourS {\ensuremath{\Upsilon{(4S)}}\xspace}
\def\FiveS {\ensuremath{\Upsilon{(5S)}}\xspace}
\def\B       {\ensuremath{B}\xspace}
\def\Bbar    {\kern 0.18em\overline{\kern -0.18em B}{}\xspace}

\def\BB      {\ensuremath{B\Bbar}\xspace}
\def\Bz      {\ensuremath{B^0}\xspace}
\def\Bzb     {\ensuremath{\Bbar^0}\xspace}
\def\BzBzb   {\ensuremath{\Bz {\kern -0.16em \Bzb}}\xspace}
\def\D       {\ensuremath{D}\xspace}
\def\Dbar    {\kern 0.18em\overline{\kern -0.18em D}{}\xspace}

\def\DD      {\ensuremath{D\Dbar}\xspace}
\def\Dz      {\ensuremath{D^0}\xspace}
\def\Dzb     {\ensuremath{\Dbar^0}\xspace}
\def\DzDzb   {\ensuremath{\Dz {\kern -0.16em \Dzb}}\xspace}

\def\Kstar   {\ensuremath{K^*}\xspace}
\def\BR       {\ensuremath{BF}\xspace}

\def\KL    {\ensuremath{K^0_{\scriptscriptstyle L}}\xspace}

\def\e#1{\ensuremath{\times10^{#1}}}

\def\makeatletter{\catcode `\@=11\relax}
\def\makeatother{ \catcode `\@=12\relax}

\makeatletter
\let\@@tau=\tau       \def\tau{\ifmath{\@@tau}}
\def\tau  {\ensuremath{\@@tau}\xspace}
\makeatother

\def\cm         {\ensuremath{\rm \,cm}\xspace}   
\def\GeV        {\ensuremath{\mathrm{\,Ge\kern -0.1em V}}\xspace}
\def\TeV        {\ensuremath{\mathrm{\,Te\kern -0.1em V}}\xspace}
\def\GeVc      {\ensuremath{{\mathrm{\,Ge\kern -0.1em V\!/}c}}\xspace}

\title{The \SuperB Project}

\ShortTitle{HQL 2010}

\author{\speaker{G. Finocchiaro}\thanks{On behalf of the \SuperB Collaboration}\\
        INFN - Laboratori Nazionali di Frascati\\
        E-mail: \email{giuseppe.finocchiaro@lnf.infn.it}}

\abstract{
The \SuperB project for a next generation asymmetric \epem flavor
factory to be built in the Rome area with a baseline luminosity of
$10^{36}\cms$ is discussed.
Some explicit examples are given to elucidate how such a 
facility can provide a uniquely sensitive probe of physics beyond the
Standard Model.
The basic accelerator concepts allowing luminosities 50-100
times larger than the existing \B factories are briefly discussed,
along with the main characteristics of the \SuperB detector.
}

\FullConference{The Xth Nicola Cabibbo International Conference on
  Heavy Quarks and Leptons,\\ 
  October 11-15, 2010\\
  Frascati (Rome) Italy}

\begin{document}

\section{Introduction}
New physics (NP) is generally expected beyond the Standard Model of
High Energy Physics (SM), although the energy scale $\Lambda$ of the
supposed new interactions is not univocally predicted.
The theoretically well-motivated possibility that $\Lambda$ is around
1\,TeV could make the NP accessible to the Large Hadron Collider (LHC).
In such a case \SuperB could study the flavor structure of NP,
measure the flavor couplings and search for still heavier mass states.
Alternatively, the NP energy scale could lie above the direct reach of
LHC. In such a scenario \SuperB can look for indirect NP signals,
understand where they may come from, and exclude regions in the
multi-dimensional parameter space of NP models up to
$\Lambda\sim10\TeV$ or more. 
In addition to probing energy scales higher than LHC via virtual
processes, \SuperB can, thanks to its clean \epem environment,
experimentally access several physics channels precluded to hadronic
machines as LHC.
Those channels include decays with neutrinos or
neutral particles in the final state, or whenever an inclusive
analysis is required, as for example $\BR(\B\to K^{(\ast)}\nunub$),
$\BR(\B\to X_s\gamma)$, and the measurement of the $|V_{ub}|$ or
$|V_{cb}|$ CKM matrix elements.

In order to achieve the sensitivity goals of the project (some
relevant examples will be discussed in the next section), a dataset of
about 75\invab  (80\e9 \BB pairs) is required, which could be
collected at the \FourS in 5 years of data taking if
$\lum=10^{36}\cms$,  assuming that the accelerator and detector
efficiency are kept at the very high levels of the \B factories. It is
important to remark that \SuperB will also produce large samples
of \D meson ($100\e9$) and \tau lepton ($70\e9$) pairs,
allowing for NP searches in the \textit{up}-type quark and lepton
sectors with unprecedented precision. The ability of \SuperB to vary
the center-of-mass energy from the charm and \tau production
threshold up to the \FiveS, and to have at least one beam with 80\%
longitudinal polarization, will further boost the physics potential of
the experiment.

As demonstrated by the several hundred papers published by the
\B-factory experiments, a wide range of important measurements
can be performed in the clean \epem environment at \FourS.
Most of  these are statistics-limited, and would therefore
improve substantially with a data sample of $75\invab$. In many cases,
large control samples can be used to further reduce systematic and
theoretical errors. Control of the theoretical errors (\eg those
related to lattice calculations) is particularly relevant to fully
exploit the statistical power of the experimental measurements.
New developments in unquenched lattice calculation techniques,
together with the expected increase in computing power,
will allow, by the time the data will be collected, to match the
\SuperB experimental precision\,\cite{ref:SuperB_Physics_WP}.
With 75\invab\ \SuperB will be able to substantially improve on the
precision of the CKM Unitarity Triangle (UT) parameters. For example,
the present error on the $\bar\rho$ and $\bar\eta$ parameters of the
UT ($\pm0.028$ and $\pm0.016$ respectively) could be
reduced\,\cite{ref:SuperB_CDR} to
$\delta\bar\rho=\pm0.0028$, $\delta\bar\eta=\pm0.0024$.
The possibility to study a very large numbers of
physical observables, and the correlations among them, is a
particularly important tool to elucidate the nature of new physics,
should deviations with respect to SM predictions be observed.

\section{\SuperB Physics Highlights}
In this section a few examples of physics channels in which the
\SuperB can give significant contributions are briefly reviewed. For
an extensive discussion, we refer the reader
to\,\cite{ref:SuperB_Physics_WP}.

\subsection{$B\to K^{(*)}\nunub$}
The rare $B\to K^{(*)}\nunub$ decays are interesting probes of new
physics from $Z^0$ penguins\,\cite{ref:BHI}.
Moreover, since the neutrinos escape the detector unmeasured,
the $b\to s+E_{miss}$ channel could be affected by exotic sources of
missing energy, such as \eg light dark
matter\,\cite{ref:LightDarkMatter_Bob}.
The $B\to K\nunub$ branching fraction, with an experimental upper
bound only a factor of three above the SM prediction, currently
provides the most stringent constraint on NP.
In the $B\to \Kstar\nunub$ decay one can also measure the longitudinal
polarization fraction $f_L(q^2)$, which
is theoretically very clean since form factor uncertainties cancel to
a large extent\,\cite{ref:Altmannshofer:2009ma}.
A detailed analysis based on the recoil technique
was performed to extrapolate
the \SuperB reach on the $B\to K^{(*)}\nunub$ decays from the
\babar\ measurement. In the recoil technique one \B meson is fully
reconstructed, yielding a
high purity sample with known kinematics, flavor and charge.
The improved \SuperB hermeticity (see later) is crucial in
background-dominated, very rare channels. In \SuperB an increase of
about 25\% with respect to \babar\ in the S/B ratio is expected.

\begin{figure}[!htb]
  \begin{center}
   \includegraphics[width=0.489\textwidth]{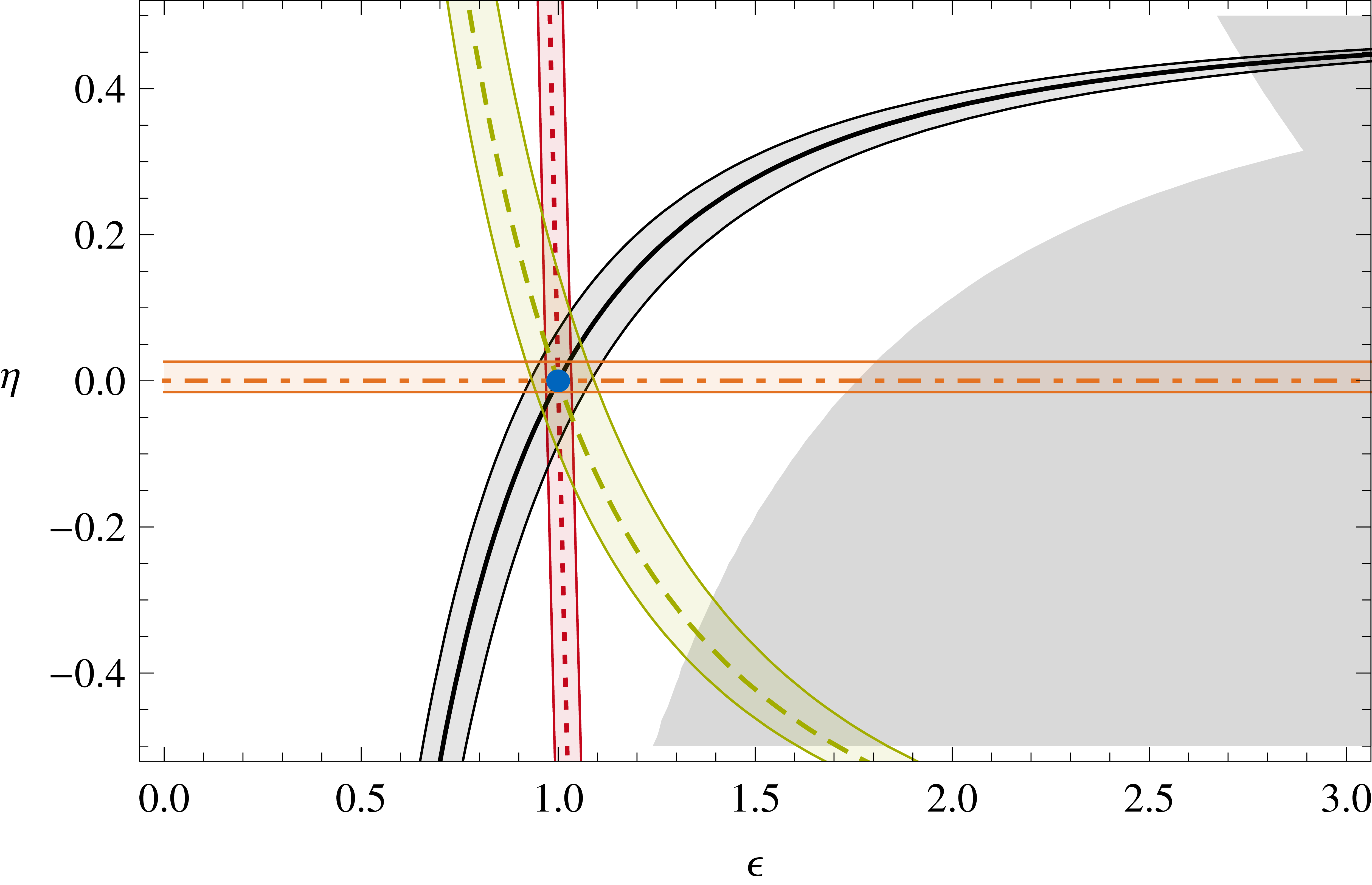}~~~~~
   \includegraphics[width=0.509\textwidth]{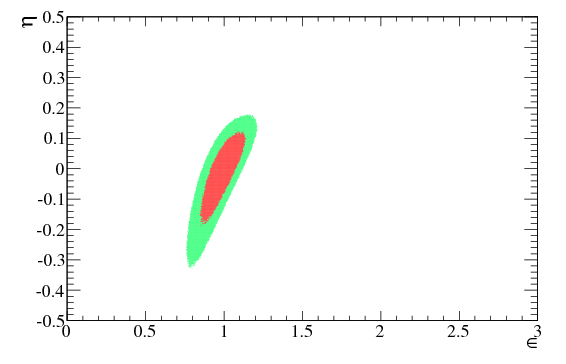}
   \caption{\label{fig:angular_constraint} Left: constraints on the
      $\epsilon-\eta$ plane, where the bands for $\BR(B\to K\nunub)$
      (green), $\BR(B\to\Kstar\nunub)$ (black),
      $f_L(q^2)$ (orange) and $\BR(B\to X_s\nunub)$ (red)
      only account for theoretical
      errors\,\cite{ref:Altmannshofer:2009ma}. The grey area is 
      excluded at 90\,\% CL by present data.
      Right: expected constraint on the
      $(\epsilon,\eta)$ plane, from the measurement of the branching
      fractions of $B \to K^{(*)} \nunub$ decays and the angular
      analysis of $B^0 \to K^{*0} \nunub$ with $75\invab$. }
  \end{center}
\end{figure}

In a general NP model $\BR(B\to K\nunub)$, $\BR(B\to
\Kstar\nunub)$, $f_L(q^2)$ and the inclusive branching fraction 
$\BR(B\to X_s\nunub)$ can be expressed in terms of the two real
parameters $\epsilon$ and $\eta$, which can be over-constrained by the
four measurements ($\epsilon=1,\eta=0$ in the SM).
Fig.\,\ref{fig:angular_constraint} shows the correlation among the
observables in the $(\epsilon,\eta)$ plane (left), and how the
constraint in the plane could be improved at \SuperB (right).

\subsection{LFV in \tau Decays}
Lepton flavor violation (LVF) in \tau decays, which is negligibly
small in the SM but is enhanced in several SM extensions, is
another very sensitive NP probe. The search for LFV has already been
actively pursued at the \B factories, which pushed the 90\,\% CL
branching fraction limits in almost 50 different decay modes down to a
few $10^{-8}$.
From an extrapolation of the \babar\ analysis,  the \SuperB
sensitivities in the ``golden'' LFV channels $\tau\to\mu\gamma$ and
$\tau\to\mu\mu\mu\gamma$ are $2.4\e{-9}$ and $2.3\e{-9}$
respectively. The ratio of branching fractions to these modes could
distinguish between the SUSY and Little Higgs
models\,\cite{ref:SuperB_CDR}.
In addition to this, the limit on $\tau\to\mu\gamma$, when combined
with constraints  on $\mu\to e\gamma$ from MEG and $\vartheta_{13}$
from accelerator and reactor neutrino experiments, can
be used to distinguish among different NP
scenarios\,\cite{ref:SuperB_CDR}.

As a key feature of its baseline design, \SuperB incorporates an
80\% polarized electron beam (polarization of the $e^+$ beam is more
difficult to realize and will be considered as an upgrade).
The angular distribution of the decay products of polarized
\tau's depends on the particular decay, and can be used to effectively
suppress backgrounds (a typical example being
$\tau\!\to\!\mu\nunub\gamma$ for the LVF decay
$\tau\!\to\!\mu\gamma$).
The polarization 
is also instrumental in improving the 
sensitivity on the \tau EDM and $g\!-\!2$.

\subsection{Charm at Threshold}
The measurement of \DzDzb oscillations has opened a new window to
search for \CP violation in charm which, if observed, would
provide unequivocal NP signals.
With 75\,\invab\ at \FourS\ \SuperB will dramatically improve (by a factor 12) the
precision in the determination of the \DzDzb mixing parameters.
The strong phase difference $\delta_f$ between the produced
\Dz and \Dzb cannot however be measured at \FourS, but only at the charm
production threshold, where the \DD pairs exhibit quantum coherence.
With 0.5\invab at the $\psi(3770)$ \SuperB can measure $\delta_f$ to
$\pm1^\circ$, several flavor-changing neural current
modes with sensitivities of the order of $10^{-8}$,
and finally strongly reduce the Dalitz-plot model uncertainty in the
$\gamma$ angle measurement.

\section{\SuperB Accelerator Highlights}
The \SuperB collider exploits a novel collision
scheme\,\cite{ref:Pantaleo,ref:Pantaleo_etal}, based on very small
beam dimensions and betatron function
at the interaction point, on large crossing and Piwinsky angle%
\footnote{The luminosity formula for \epem beams colliding with an horizontal
crossing angle $\vartheta$ can be expressed\,\cite{ref:Pantaleo_etal} in terms
of the vertical tune shift parameter $\xi_y$, the vertical beta
function at the IP $\beta_y$ and the number of particles per bunch $N$
(proportional to the beam current) as
$\lum\propto{N\xi_y}/{\beta_y}$, with
$\xi_y\propto{N\beta_y}/({\sigma_x\sigma_y\sqrt{1+\phi^2}})$. The
Piwinsky angle $\phi\simeq\vartheta\sigma_x/\sigma_y$.}
and on the ``crab waist'' scheme. This approach allows
to reach the required luminosity of $10^{36}\cms$ and at the same time
overcome the difficulties of early super \epem collider
designs, most notably very high beam currents and very short bunch
lengths. The wall-plug power and the beam-related background rates in
the detector are therefore kept within affordable levels\,\cite{ref:SuperB_Collider_WP}.

The crab waist transformation consists in moving the waist of each
beam onto the axis of the other beam with a pair of sextupole up- and
down-stream the IP. In this way all particles from both beams collide
in the minimum $\beta_y^*$ region, with a net luminosity gain.
Moreover (and most significantly) the $x/y$ betatron resonances
are naturally suppressed.
The principle of the innovative IR design sketched above has been
experimentally demonstrated at the Frascati Da$\phi$NE
collider\,\cite{ref:dafnetest}. Very importantly, this test also
validated the simulations used to calculate the IR optics.

Several other accelerator-physics ideas critical to the success of the
\SuperB project have also been realized in practice, either in past 
colliders or as part of accelerator R\&D activities such as those for
the Linear Collider: KEK-B and PEP-II have worked very well with
asymmetric interaction regions, storing 2-3A of beam currents and
performing continuous injection with live detectors;
polarized beams have been successfully produced at the SLC, and spin
manipulation tests have been performed at Novosibirsk;
finally, ultra-low emittance lattices were tested for the ILC damping
rings. This long list of achievements, and the fact that the
luminosity of $10^{36}\cms$ is not a
singularity in the parameter space but can be obtained with different
settings of the accelerator parameters, as \eg the $x$ and $y$
emittances and $\beta_y^*$ in both rings, the vertical tune shift, the
beam currents, adds confidence that this unprecedented luminosity can
indeed be obtained in practice.

The \SuperB design is based on recycling as much as possible existing
PEP-II hardware, with a significant reduction of costs. The optimal beam
energy choice for accelerator design, including polarization, is
4.18\GeV positron beam on 6.71\GeV electron beam.

The low currents, ultra-small emittance approach has been
adopted recently also by the KEKB accelerator team, which defined
a new set of parameters very similar to that of the Italian \SuperB.

\section{\SuperB Detector Highlights}
Most of the general requirements for the \SuperB detector are common
to those of the present \B factories, including large solid angle
coverage, good particle identification (PID) capabilities over a
wide momentum range ($\pi/K$ separation to over 4\GeVc),
measurement of the relative decay times of the \B mesons,
good resolution on the momentum of charged tracks and on the photon
energy, particularly in the sub-\GeV part of the spectrum, relevant at
the \FourS environment.
The \SuperB detector concept is therefore based on the \babar
detector, with the modifications required to operate at a much higher
luminosity (and luminosity-scaling background rates), and with a
reduced center-of-mass boost\,\cite{ref:SuperB_Detector_WP}.

The \babar detector is composed by a tracking system -- a
five layer double-sided silicon strip vertex tracker (SVT) and a 40 layer
drift chamber (DCH) immersed in a 1.5\,T magnetic field -- a Cherenkov
detector with fused silica bar radiators (DIRC), a homogeneous
electromagnetic calorimeter made of CsI(Tl) crystals (EMC), and a
detector for muon identification and \KL detection (IFR) realized
instrumenting the iron flux return with resistive plate chambers
and limited streamer tubes.
\SuperB is designed to reuse a number of \babar components: the DIRC
quartz bars, the CsI(Tl) crystals of the barrel EMC, the flux-return steel,
the superconducting coil.

\begin{figure*}[hbtp]
  \begin{center}
    \includegraphics[width=\textwidth]{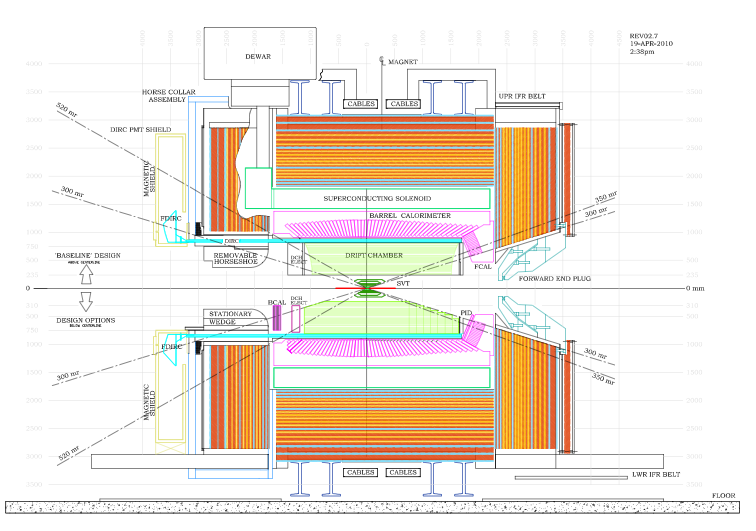}
    \caption{Concept for the \SuperB detector.  The upper half shows
      the baseline concept, and the bottom half adds a number of
      optional detector configurations.}
    \label{fig:det:superb}
  \end{center}
\end{figure*}

The center-of-mass boost at \SuperB is smaller than in \babar
($\beta\gamma=0.24$ vs. $0.56$). While this effectively improves the
angular coverage of the detector, it also reduces the $\Delta z$
separation of the decay vertices. 
The $\Delta t$ sensitivity in time-dependent measurements is
maintained by improving the vertex resolution: the \SuperB vertex
detector replicates the five-layer \babar SVT, but exploits the
reduced dimensions of the beam pipe made possible by the
ultra-low emittance \SuperB beams to add a very thin and precise
measurement layer at a radius of only 1.5\cm.
The baseline technology for this ``Layer0'' uses short double-sided
silicon strip detectors (``striplets''), while other options are
being considered as possible upgrades.
The \SuperB DCH concept is derived from the \babar one,
with several improvements: the mechanical structure entirely in
Carbon-Fiber composite, resulting in a 4-fold reduction of the
endplate material; the gas mixture and the wire cell layout optimized
to minimize the multiple-scattering contribution to momentum
resolution; the readout electronics redesigned to cope with
the higher trigger rate, and minimize the FEE material.
The hadron PID system will use the radiator quartz bars of the
\babar DIRC, read-out by fast multi-anode PMTs, and with the imaging
region considerably reduced in size to improve performance and reduce
the impact of backgrounds. The forward EMC will feature cerium-doped
LYSO crystals, which have a much
shorter scintillation time constant, a smaller Moli\`ere radius and better
radiation hardness than the current CsI(Tl) crystals, for
reduced sensitivity to beam backgrounds and better position
resolution. The thickness of the flux-return iron will be increased with additional
absorber to bring to about 7 the number of interaction lengths for
muons, while the gas detectors will be replaced by extruded plastic
scintillator bars to cope with the expected background rates.
Finally, the Collaboration is considering to improve the detector
hermeticity by inserting a ``veto-quality'' lead-scintillator EMC
calorimeter in the backward direction, and to add a particle identification
device in front of the forward calorimeter. Both detectors would
improve the sensitivity to rare modes such as $\B\to
K^{(\ast)}\nunub$ and $\B\to\tau\nu$.
A concept of the \SuperB detector is shown in Fig.\,\ref{fig:det:superb}.

\section{Summary}
We concisely presented the \SuperB physics program, discussing some
examples of key measurements possible at \FourS and $\psi(3770)$ 
center-of-mass energies, and the importance of incorporating
longitudinal beam polarization in the design. The main features of the
accelerator and detector projects were briefly described.
An exhaustive discussion of all \SuperB aspects can be found
in\,\cite{ref:SuperB_Physics_WP,ref:SuperB_Collider_WP,ref:SuperB_Detector_WP}.

\end{document}